# Linear Coupling:
# An Ultimate Unification of Gradient and Mirror Descent[*]

Zeyuan Allen-Zhu
zeyuan@csail.mit.edu
Institute for Advanced Study

Lorenzo Orecchia
orecchia@bu.edu
Boston University

November 7, 2016

**Abstract**

First-order methods play a central role in large-scale machine learning. Even though many variations exist, each suited to a particular problem, almost all such methods fundamentally rely on two types of algorithmic steps: gradient descent, which yields primal progress, and mirror descent, which yields dual progress.

We observe that the performances of gradient and mirror descent are complementary, so that faster algorithms can be designed by *linearly coupling* the two. We show how to reconstruct Nesterov's accelerated gradient methods using linear coupling, which gives a cleaner interpretation than Nesterov's original proofs. We also discuss the power of linear coupling by extending it to many other settings that Nesterov's methods cannot apply to.

## 1 Introduction

The study of fast iterative methods for approximately solving convex problems is a central research focus in Machine Learning, Combinatorial Optimizations and many other areas of Computer Science and Mathematics. For large-scale programs, first-order iterative methods are usually the methods of choice due to their cheap and often highly parallelizable iterations.

First-order methods access the target optimization problem $\min_{x \in Q} f(x)$ in a black-box fashion: the algorithm queries a point $y \in Q$ at every iteration and receives the pair $\big(f(y), \nabla f(y)\big)$. [1] The complexity of a first-order method is usually measured in the number of queries necessary to produce an additive $\varepsilon$-approximate minimizer. First-order methods have recently experienced a renaissance in the design of fast algorithms for fundamental computer science problems, varying from discrete ones such as maximum flow problems [20], to continuous ones such as empirical risk minimization [39].

Despite the myriad of applications, first-order methods with provable convergence guarantees can be mostly classified as instantiations of two fundamental algorithmic ideas: *gradient descent* and the *mirror descent*.[2] We argue that gradient descent takes a fundamentally primal approach, while mirror descent follows a complementary dual approach. In our main result, we show how these two approaches blend in a natural manner to yield a new and simple accelerated gradient method for smooth convex optimization problems, as well as lead to other applications where the classical accelerated gradient methods do not apply.

---

[*]The authors would like to thank Silvio Micali for listening to our work and suggesting the name "linear coupling".

[1]Here, variable $x$ is constrained to lie in a convex set $Q \subseteq \mathbb{R}^n$, which is known as the *constraint set* of the problem.

[2]We emphasize here that these two terms are sometimes used ambiguosly in the literature; in this paper, we attempt to stick as close as possible to the conventions of the optimization community and in particular in the textbooks [9, 26] with one exception: we extend the definition of gradient descent to non-Euclidean norms in a natural way, following [18].

## 1.1 Understanding First-Order Methods: Gradient Descent and Mirror Descent

We now provide high-level descriptions of gradient and mirror descent. While this material is classical, our intuitive presentation of these ideas forms the basis for our main result in the subsequent sections. For a more detailed survey, we recommend the textbooks [9, 26].

Consider for simplicity the unconstrained minimization (i.e. $Q = \mathbb{R}^n$), but, as we will see in Section 2, the same intuition and a similar analysis extend to the constrained or even the proximal case. We use generic norms $\|\cdot\|$ and their duals $\|\cdot\|_*$. At a first reading, they can be both replaced with the Euclidean norm $\|\cdot\|_2$.

### 1.1.1 Primal Approach: Gradient Descent

A natural approach to iterative optimization is to decrease the objective function as much as possible at every iteration. To formalize the effectiveness of this idea, one usually introduces a smoothness assumption on the objective $f(x)$. Specifically, recall that an $L$-smooth function $f$ satisfies $\|\nabla f(x) - \nabla f(y)\|_* \leq L\|x - y\|$ for every $x, y$. Such a smoothness condition yields a global quadratic upper bound on the function around a query point $x$:

$$\forall y, \quad f(y) \leq f(x) + \langle \nabla f(x), y - x \rangle + \frac{L}{2}\|y - x\|^2 \ . \tag{1.1}$$

Gradient-descent algorithms exploit this bound by taking a step that maximizes the guaranteed objective decrease (i.e., the primal progress) $f(x_k) - f(x_{k+1})$ at every iteration $k$. More precisely,

$$x_{k+1} \leftarrow \arg\min_y \left\{ \frac{L}{2}\|y - x_k\|^2 + \langle \nabla f(x), y - x_k \rangle \right\} \ .$$

Notice that here $\|\cdot\|$ is a generic norm. When this is the Euclidean $\ell_2$-norm, the step takes the familiar additive form $x_{k+1} = x_k - \frac{1}{L}\nabla f(x_k)$. However, in other cases, e.g., for the non-Euclidean $\ell_1$ or $\ell_\infty$ norms, the update step will not follow the direction of the gradient $\nabla f(x_k)$ (see for instance [18, 27]).

Under the smoothness assumption above, the magnitude of this primal progress is at least

$$f(x_k) - f(x_{k+1}) \geq \frac{1}{2L}\|\nabla f(x_k)\|_*^2 \ . \tag{1.2}$$

In general, this quantity will be larger when the gradient $\nabla f(x_k)$ has large norm. Classical convergence analysis of gradient descent usually combines (1.2) with a basic convexity argument to relate $f(x_k) - f(x^*)$ and $\|\nabla f(x_k)\|_*$: that is, $f(x_k) - f(x^*) \leq \|\nabla f(x_k)\|_*\|x_k - x^*\|$. For $L$-smooth objectives, the final bound shows that gradient descent converges in $O\left(\frac{L}{\varepsilon}\right)$ iterations [26].

The limitation of gradient descent is that it does not make any attempt to construct a good lower bound to the optimum value $f(x^*)$. It essentially ignores the dual problem. In the next subsection, we review mirror descent, a method that focuses completely on the dual side.

### 1.1.2 Dual Approach: Mirror Descent

Mirror-descent methods (see for instance [9, 12, 24, 28, 44]) tackle the dual problem by constructing lower bounds to the optimum. Recall that each queried gradient $\nabla f(x)$ can be viewed as a hyperplane lower bounding the objective $f$: that is, $f(u) \geq f(x) + \langle \nabla f(x), u-x \rangle$ for all $u$. Mirror-descent methods attempt to carefully construct a convex combination of these hyperplanes in order to yield even a stronger lower bound. Formally, suppose one has queried points $x_0, \ldots, x_{k-1}$, then we form a linear combination of the $k$ hyperplanes and obtain[3]

$$\forall u, \quad f(u) \geq \frac{1}{k} \sum_{t=0}^{k-1} f(x_t) + \frac{1}{k} \sum_{t=0}^{k-1} \langle \nabla f(x_t), u - x_t \rangle \ . \tag{1.3}$$

On the upper bound side, we consider a simple choice $\overline{x} = \frac{1}{k} \sum_{t=0}^{k-1} x_t$, i.e., the mean of the queried points. By straightforward convexity argument, we have $f(\overline{x}) \leq \frac{1}{k} \sum_{t=0}^{k-1} f(x_t)$. As a result, the distance between $f(\overline{x})$ and $f(u)$ for any arbitrary $u$ can be upper bounded using (1.3):

$$\forall u, \quad f(\overline{x}) - f(u) \leq \frac{1}{k} \sum_{t=0}^{k-1} \langle \nabla f(x_t), x_t - u \rangle \stackrel{\text{def}}{=} R_k(u) \ . \tag{1.4}$$

Borrowing terminology from online learning, the right hand side $R_k(u)$ is known as the *regret* of the sequence $(x_t)_{t=0}^{k-1}$ with respect to point $u$. Now, consider a regularized version $\widetilde{R}_k(u)$ of the regret

$$\widetilde{R}_k(u) \stackrel{\text{def}}{=} \frac{1}{k} \cdot \left( -\frac{w(u)}{\alpha} + \sum_{t=0}^{k-1} \langle \nabla f(x_t), x_t - u \rangle \right) \ ,$$

where $\alpha > 0$ is a trade-off parameter and $w(\cdot)$ is some regularizer that is usually strongly convex. Then, mirror-descent methods choose the next iterate $x_k$ by minimizing the maximum regularized regret at the next iteration: that is, choose $x_k \leftarrow \arg\max_u \widetilde{R}_k(u)$. This update rule can be shown to successfully drive $\max_u \widetilde{R}_k(u)$ down as $k$ increases, and thus the right hand side of (1.4) decreases as $k$ increases. This can be made into a rigorous analysis and show that mirror descent converges in $T = O(\rho^2/\varepsilon^2)$ iterations. Here, $\rho^2$ is the average value of $\|\nabla f(x_k)\|_*^2$ across the iterations.

To sum up, the smaller the queried gradients are (i.e. the smaller $\|\nabla f(x_k)\|_*$ is), the tighter the lower bound (1.3) becomes, and therefore the fewer iterations are needed for mirror descent to converge. (Note that the above mirror-descent analysis can also be used to derive the $1/\varepsilon$ convergence rate on smooth objectives similar to that in gradient descent [11]; since this adaption is not needed in our paper, we omit the details.)

**Remarks.** Mirror descent admits several different algorithmic implementations, such as *Nemirovski's mirror descent* [24] and *Nesterov's dual averaging* [28].[4] Results based on one implementation can usually be transformed into another with some efforts. In this paper, we adopt Nemirovski's mirror descent as our choice of mirror descent, see Section 2.2.

One may occasionally find analyses that do not immediately fall into the above two categories. To name a few, solely using mirror descent and dual lower bounds, one can also obtain a convergence rate $1/\varepsilon$ for smooth objectives similar to that in gradient descent [11]. Conversely, one can deduce the mirror-descent guarantee by applying gradient descent on a dual objective (see Appendix A.3). Shamir and Zhang [40] obtained an algorithm that converges slightly slower than mirror descent, but has an error guarantee on the last iterate, rather than the average history.

---

[3]For simplicity, we choose uniform weights here. For the purpose of proving convergence results, the weights of individual hyperplanes are typically uniform or only dependent on $k$.

[4]Other update rules can be viewed as specializations or generalizations of the mentioned implementations. For instance, the follow-the-regularized-leader (FTRL) step is a generalization of Nesterov's dual averaging step where the regularizers are can be adaptively selected (see [23]).

## 1.2 Our Conceptual Question

Following this high level description of gradient and mirror descent, it is useful to pause and observe the complementary nature of the two procedures. Gradient descent relies on primal progress, uses local steps and makes faster progress when the norms of the queried gradients $\|\nabla f(x_k)\|$ are *large*. In contrast, mirror descent works by ensuring dual progress, uses global steps and converges faster when the norms of the queried gradients $\|\nabla f(x_k)\|$ are *small*.

This interpretation immediately leads to the question that inspires our work:

*Can Gradient Descent and Mirror Descent be combined to obtain faster first-order methods?*

In this paper, we initiate the formal study of this key conceptual question, and propose a *linear coupling* framework. To properly discuss our framework, we choose to mostly focus in the context of convex smooth minimization, and show how to reconstruct Nesterov's accelerated gradient methods using linear coupling. We also discuss the power of our framework by extending it to many other settings beyond Nesterov's original scope.

## 1.3 Accelerated Gradient Method Via Linear Coupling

In the seminal work [25, 26], Nesterov designed an accelerated gradient method for $L$-smooth functions with respect to $\ell_2$ norms, and it performs quadratically faster than gradient descent — requiring $\Omega(L/\varepsilon)^{0.5}$ rather than $\Omega(L/\varepsilon)$ iterations. This is asymptotically tight [26]. Later in 2005, Nesterov generalizes his method to allow non-Euclidean norms in the definition of smoothness [27]. All these versions of methods are referred to as *accelerated gradient methods*, or sometimes as Nesterov's accelerated methods.

Although accelerated gradient methods have been widely applied (to mention a few, see [38, 39] for regularized optimizations, [19, 30] for composite optimization, [29] for cubic regularization, [31] for universal method, and [20] for an application on maxflow), they are often regarded as "analytical tricks" [17] because their convergence analyses are somewhat complicated and lack of intuitions.

In this paper, we provide a simple, alternative, but **complete** version of the accelerated gradient method. Here, by "complete" we mean our method works for any norm, and for both the constrained and unconstrained case.[5] Our key observation is to construct two sequences of updates: one sequence of gradient-descent updates and one sequence of mirror-descent updates.

**Thought Experiment.** Consider $f(x)$ that is unconstrained and $L$-smooth. For sake of demonstrating the idea, suppose $\|\nabla f(x)\|_2$, the norm of the observed gradient, is *either* always $\geq K$, or always $\leq K$, where the cut-off value $K$ is determined later. Under such "wishful assumption", we propose the following algorithm: if $\|\nabla f(x)\|_2$ is always $\geq K$, we perform $T$ gradient-descent steps; otherwise we perform $T$ mirror-descent steps.

To analyze such an algorithm, suppose without loss of generality we start with some point $x_0$ whose objective distance $f(x_0) - f(x^*)$ is at most $2\varepsilon$, and we want to find some $x$ so that $f(x) - f(x^*) \leq \varepsilon$.[6] If $T$ gradient-descent steps are performed, the objective decreases by at least $\frac{\|\nabla f(\cdot)\|_2^2}{2L} \geq \frac{K^2}{2L}$ per step according to (1.2), and we only need $T \geq \Omega(\frac{\varepsilon L}{K^2})$ steps to achieve an $\varepsilon$ accuracy. If $T$ mirror-descent steps are performed, we need $T \geq \Omega(\frac{K^2}{\varepsilon^2})$ steps according to the mirror-descent convergence. In sum, we need $T \geq \Omega\big(\max\big\{\frac{\varepsilon L}{K^2}, \frac{K^2}{\varepsilon^2}\big\}\big)$ steps to converge to an $\varepsilon$-minimizer. Setting $K$ to be the "magic number" to balance the two terms, we only need $T \geq \Omega\big(\frac{L}{\varepsilon}\big)^{1/2}$ iterations as desired.

---

[5]Some authors have regarded the result in [26] as "momentum analysis" [32, 41] or "ball method" [10]. These analyses only apply to Euclidean spaces. We point out the importance of allowing non-Euclidean norms in Appendix A.1. In addition, our proof in this paper extends naturally to the proximal version of first-order methods, but for simplicity, we choose to include only the constrained version.

**Towards an Actual Proof.** To turn our thought experiment into an actual proof, we face the following obstacles. Although gradient-descent steps always decrease the objective, mirror-descent steps may sometimes increase the objective, cancelling the effect of the gradient descent. On the other hand, the mirror-descent steps are *only* useful when a large number of iterations are performed in a row; if any gradient-descent step stands in the middle, the convergence is destroyed.

For this reason, it is natural to design an algorithm that, in every single iteration $k$, performs *both* a gradient and a mirror descent step, and somehow ensure that the two steps are coupled together. However, the following additional difficulty arises: if from some starting point $x_k$, the gradient-descent step instructs us to go to $y_k$, while the mirror-descent step instructs us to go to $z_k$, then how do we continue? Do we look at the gradient at $\nabla f(y_k)$ or $\nabla f(z_k)$ in the next iteration?

This problem is implicitly solved by Nesterov using the following simple idea[7]: in the $k$-th iteration, we choose a linear combination $x_{k+1} \leftarrow \tau z_k + (1-\tau) y_k$, and use this same gradient $\nabla f(x_{k+1})$ to continue the gradient and mirror steps of the next iteration. Whenever $\tau$ is carefully chosen (just like the "magic number" $K$), the two descent sequences provide a coupled bound on the error guarantee, and we recover the same convergence as [27].

**Roadmap.** We review the key lemmas of gradient and mirror descent in Section 2. We propose a simple method with fixed step length to recover Nesterov's accelerated methods for the unconstrained case in Section 3, and generalize it to the full-setting in Section 4. We discuss several important applications of linear coupling that Nesterov's original methods do not solve in Section 5.

## 2 Key Lemmas of Gradient and Mirror Descent

### 2.1 Review of Gradient Descent

Consider a function $f(x)$ that is convex and differentiable on a closed convex set $Q \subseteq \mathbb{R}^n$, and assume that $f$ is *L-smooth* with respect to $\|\cdot\|$, that is, for every $x, y \in Q$, it satisfies $\|\nabla f(x) - \nabla f(y)\|_* \leq L\|x - y\|$. Here, $\|\cdot\|_*$ is the dual norm of $\|\cdot\|$.[8]

**Definition 2.1.** *For any $x \in Q$, the* gradient (descent) step *(with step length $\frac{1}{L}$) is*

$$\widetilde{x} = \mathsf{Grad}(x) \stackrel{\text{def}}{=} \arg\min_{y \in Q} \Big\{\frac{L}{2}\|y - x\|^2 + \langle \nabla f(x), y - x\rangle\Big\}$$

*and we let* $\mathsf{Prog}(x) \stackrel{\text{def}}{=} -\min_{y \in Q}\big\{\frac{L}{2}\|y-x\|^2 + \langle \nabla f(x), y-x\rangle\big\} \geq 0$.

In particular, when $\|\cdot\| = \|\cdot\|_2$ is the $\ell_2$-norm and $Q = \mathbb{R}^n$ is unconstrained, the gradient step can be simplified as $\mathsf{Grad}(x) = x - \frac{1}{L}\nabla f(x)$. Or, slightly more generally, when $\|\cdot\| = \|\cdot\|_2$ is the $\ell_2$-norm but $Q$ may be constrained, we have $\mathsf{Grad}(x) = x - \frac{1}{L}g_Q(x)$ where $g_Q(x)$ is the gradient mapping of $f$ at $x$ (see Chapter 2.2.3 of [26]).

---

[6]For all first-order methods, the heaviest computation always happens in this $2\varepsilon$ to $\varepsilon$ process.

[7]We wish to point out that Nesterov has phrased his method differently from ours, and little is known on why this linear combination is needed from his proof, except for being used as an algebraic trick to cancel specific terms.

[8]$\|\xi\|_* \stackrel{\text{def}}{=} \max\{\langle \xi, x\rangle : \|x\| \leq 1\}$. For instance, $\ell_p$ norm is dual to $\ell_q$ norm if $\frac{1}{p} + \frac{1}{q} = 1$.

The classical theory on smooth convex programming gives rise to the following lower bound on the amount of objective decrease (proved in Appendix B for completeness):

$$f(\mathsf{Grad}(x)) \leq f(x) - \mathsf{Prog}(x) \tag{2.1}$$

$$\text{or in the special case when } Q = \mathbb{R}^n \quad f(\mathsf{Grad}(x)) \leq f(x) - \frac{1}{2L}\|\nabla f(x)\|_*^2 \ .$$

From the above descent guarantee, one can deduce the convergence rate of gradient descent. For instance, if $\|\cdot\| = \|\cdot\|_2$ is the Euclidean norm, and if gradient step $x_{k+1} = \mathsf{Grad}(x_k)$ is applied $T$ times, we obtain the following convergence guarantee (see [26])

$$f(x_T) - f(x^*) \leq O\Big(\frac{L\|x_0 - x^*\|_2^2}{T}\Big) \quad \text{or equivalently} \quad T \geq \Omega\Big(\frac{L\|x_0 - x^*\|_2^2}{\varepsilon}\Big) \Rightarrow f(x_T) - f(x^*) \leq \varepsilon \ .$$

Here, $x^*$ is any minimizer of $f(x)$. If $\|\cdot\|$ is a general norm, but $Q = \mathbb{R}^n$ is unconstrained, the above convergent rate becomes $f(x_T) - f(x^*) \leq O(\frac{LR^2}{T})$, where $R = \max_{x:f(x)\leq f(x_0)} \|x - x^*\|$. We provide the proof of this later case in Appendix B because it is less known and we cannot find it in the optimization literature.

Note that, we are unaware of any universal convergence proof for both the general norm and the unconstrained case. As we shall see later in Section 4, this convergence rate can be improved by accelerated gradient methods, even for the general norm $\|\cdot\|$ and the constrained case.

## 2.2 Review of Mirror Descent

Consider some function $f(x)$ that is convex on a closed convex set $Q \subseteq \mathbb{R}^n$, and assume that $f$ is $\rho$-*Lipschitz continuous* with respect to norm $\|\cdot\|$, that is, for every $x, y \in Q$, it satisfies $|f(x) - f(y)| \leq \rho\|x - y\|$. This is equivalent to saying that $f$ admits a subgradient $\partial f(x)$ at every point $x \in Q$, and satisfies $\|\partial f(x)\|_* \leq \rho$. (Recall that $\partial f(x) = \nabla f(x)$ if $f$ is differentiable.)

Mirror descent requires one to choose a regularizer (also referred to as a distance generating function):

**Definition 2.2.** *We say that $w \colon Q \to \mathbb{R}$ is a* distance generating function (DGF), *if $w$ is 1-strongly convex with respect to $\|\cdot\|$, or in symbols, $\forall x \in Q \setminus \partial Q, \forall y \in Q$: $w(y) \geq w(x) + \langle \nabla w(x), y - x\rangle + \frac{1}{2}\|x - y\|^2$. Accordingly, the* Bregman divergence *is given as*

$$V_x(y) \stackrel{\mathrm{def}}{=} w(y) - \langle \nabla w(x), y - x\rangle - w(x) \quad \forall x \in Q \setminus \partial Q, \forall y \in Q \ .$$

*The property of DGF ensures that $V_x(x) = 0$ and $V_x(y) \geq \frac{1}{2}\|x - y\|^2 \geq 0$.*

Common examples of DGFs include (i) $w(y) = \frac{1}{2}\|y\|_2^2$, which is strongly convex with respect to the $\ell_2$-norm over every $Q$, and the corresponding $V_x(y) = \frac{1}{2}\|x - y\|_2^2$, and (ii) the entropy function $w(y) = \sum_i y_i \log y_i$, which is strongly convex with respect to the $\ell_1$-norm over any $Q \subseteq \Delta \stackrel{\mathrm{def}}{=} \{x \geq 0 : \mathbb{1}^T x = 1\}$. and the corresponding $V_x(y) = \sum_i y_i \log(y_i/x_i) \geq \frac{1}{2}\|x - y\|_1^2$.

**Definition 2.3.** *The* mirror (descent) step *with step length $\alpha$ can be described as*

$$\widetilde{x} = \mathsf{Mirr}_x(\alpha \cdot \partial f(x)) \quad where \quad \mathsf{Mirr}_x(\xi) \stackrel{\mathrm{def}}{=} \arg\min_{y \in Q} \big\{V_x(y) + \langle \xi, y - x\rangle\big\} \ .$$

Mirror descent's core lemma is the following inequality (proved in Appendix B for completeness):

If $x_{k+1} = \mathsf{Mirr}_{x_k}(\alpha \cdot \partial f(x_k))$, then

$$\forall u \in Q, \quad \alpha(f(x_k) - f(u)) \leq \alpha \langle \partial f(x_k), x_k - u \rangle \leq \frac{\alpha^2}{2} \|\partial f(x_k)\|_*^2 + V_{x_k}(u) - V_{x_{k+1}}(u) \quad (2.2)$$

The term $\langle \partial f(x_k), x_k - u \rangle$ features prominently in online optimization, and is known as the *regret* at iteration $k$ with respect to $u$ (see Appendix A.2 for the folklore relationship between mirror descent and regret minimization). It is not hard to see that, telescoping (2.2) for $k = 0, \ldots, T-1$, setting $\overline{x} \stackrel{\text{def}}{=} \frac{1}{T} \sum_{k=0}^{T-1} x_k$ to be the average of the $x_k$'s, and choosing $u = x^*$, we have

$$\alpha T(f(\overline{x}) - f(x^*)) \leq \sum_{k=0}^{T-1} \alpha \langle \partial f(x_k), x_k - x^* \rangle \leq \frac{\alpha^2}{2} \sum_{k=0}^{T-1} \|\partial f(x_k)\|_*^2 + V_{x_0}(x^*) - V_{x_T}(x^*) \ . \quad (2.3)$$

Finally, letting $\Theta$ be any upper bound on $V_{x_0}(x^*)$ (recall $\Theta = \frac{1}{2}\|x_0 - x^*\|_2^2$ when $\|\cdot\|$ is the Euclidean norm), and $\alpha = \frac{\sqrt{2\Theta}}{\rho \cdot \sqrt{T}}$ be the step length, inequality (2.2) can be re-written as

$$f(\overline{x}) - f(x^*) \leq \frac{\sqrt{2\Theta} \cdot \rho}{\sqrt{T}} \quad \text{or equivalently} \quad T \geq \frac{2\Theta \cdot \rho^2}{\varepsilon^2} \Rightarrow f(\overline{x}) - f(x^*) \leq \varepsilon \ . \quad (2.4)$$

**Remark.** While their analyses share some similarities, mirror and gradient steps are often very different. For example, if the optimization problem is over the simplex with $\ell_1$ norm, then gradient step gives $x' \leftarrow \arg\min_y \{\frac{1}{2}\|y-x\|_1^2 + \alpha \langle \nabla f(x), y-x \rangle\}$, while the mirror step with entropy regularizer gives $x' \leftarrow \arg\min_y \{\sum_i y_i \log(y_i/x_i) + \alpha \langle \nabla f(x), y-x \rangle\}$. We point out in Appendix A.1 that non-Euclidean norms are very important for certain applications.

In the special case of $w(x) = \frac{1}{2}\|x\|_2^2$ and $\|\cdot\|$ is $\ell_2$-norm, gradient and mirror steps are indistinguishable from each other. However, as we have discussed earlier, these two update rules are often equipped with very different convergence analyses, even if they 'look the same'.

## 3 Warm-Up Method with Fixed Step Length

Consider the same setting as Section 2.1: that is, $f(x)$ is convex and differentiable on its domain $Q$, and is $L$-smooth with respect to some norm $\|\cdot\|$. (Note that $f(x)$ may not have a good Lipschitz continuity parameter $\rho$, but we do not need such a property.) In this section, we focus on the unconstrained case $Q = \mathbb{R}^n$, and combine gradient and mirror descent to produce a very simple accelerated method. We explain this method first because it avoids the mysterious choices of step lengths as in the full setting, and carries our conceptual message in a very clean way.

Design an algorithm that, in every step $k$, performs *both* a gradient and a mirror step, and ensures that the two steps are linearly coupled. More specifically, starting from $x_0 = y_0 = z_0$, in each iteration $k = 0, 1, \ldots, T-1$, we first define $x_{k+1} \leftarrow \tau z_k + (1-\tau)y_k$ and then
- perform a gradient step $y_{k+1} \leftarrow \mathsf{Grad}(x_{k+1})$, and
- perform a mirror step $z_{k+1} \leftarrow \mathsf{Mirr}_{z_k}(\alpha \nabla f(x_{k+1}))$.[9]

Above, $\alpha$ is the (fixed) step length of the mirror step, while $\tau$ is the parameter controlling the coupling rate. The choices of $\alpha$ and $\tau$ will become clear at the end of this section, but from a high level,
- $\alpha$ will be determined from the mirror-descent analysis, similar to that in (2.3), and

---
[9]Here, the mirror step $\mathsf{Mirr}$ is defined by specifying any DGF $w(\cdot)$ that is 1-strongly convex over $Q$.

- $\tau$ will be determined as the best parameter to balance the gradient and mirror steps, similar to the "magic number" $K$ in our thought experiment discussed in Section 1.3.

Classical gradient-descent and mirror-descent analyses immediately imply the following:

**Lemma 3.1.** *For every $u \in Q = \mathbb{R}^n$,*

$$\alpha \langle \nabla f(x_{k+1}), z_k - u \rangle \overset{\text{①}}{\leq} \frac{\alpha^2}{2} \|\nabla f(x_{k+1})\|_*^2 + V_{z_k}(u) - V_{z_{k+1}}(u)$$

$$\overset{\text{②}}{\leq} \alpha^2 L \big( f(x_{k+1}) - f(y_{k+1}) \big) + V_{z_k}(u) - V_{z_{k+1}}(u) \ . \tag{3.1}$$

*Proof.* To deduce ①, we note that our mirror step $z_{k+1} = \mathsf{Mirr}_{z_k}(\alpha \nabla f(x_{k+1}))$ is essentially identical to that of $x_{k+1} = \mathsf{Mirr}_{x_k}(\alpha \nabla f(x_k))$ in (2.2), with only changes of variable names. Therefore, inequality ① is a simple copy-and-paste from (2.2) after changing the variable names (see the proof of (2.2) for details). The second inequality ② is from the gradient step guarantee $f(x_{k+1}) - f(y_{k+1}) \geq \frac{1}{2L} \|\nabla f(x_{k+1})\|_*^2$ in (2.1). □

One can immediately see from Lemma 3.1 that, although the mirror step introduces an error $\frac{\alpha^2}{2} \|\nabla f(x_{k+1})\|_*^2$, this error is proportional to the amount of the gradient-step progress $f(x_{k+1}) - f(y_{k+1})$. This captures the observation we stated in the introduction: if $\|\nabla f(x_{k+1})\|_*$ is large, we can make a large gradient step, or if $\|\nabla f(x_{k+1})\|_*$ is small, the mirror step suffers from a small loss.

If we choose $\tau = 1$ or equivalently $x_{k+1} = z_k$, the left hand side of inequality (3.1) becomes $\langle \nabla f(x_{k+1}), x_{k+1} - u \rangle$, the regret at iteration $x_{k+1}$. In such a case we wish to telescope it for all iterations $k$ in the spirit of mirror descent (see (2.3)). However, we face the problem that the terms $f(x_{k+1}) - f(y_{k+1})$ do not telescope. [10] On the other hand, if we choose $\tau = 0$ or equivalently $x_{k+1} = y_k$, then the terms $f(x_{k+1}) - f(y_{k+1}) = f(y_k) - f(y_{k+1})$ telescope, but the left hand side of (3.1) is no longer the regret. [11]

To overcome this issue, we use linear coupling. We compute and upper bound the difference between the left hand side of (3.1) and the actual "regret":

$$\alpha \langle \nabla f(x_{k+1}), x_{k+1} - u \rangle - \alpha \langle \nabla f(x_{k+1}), z_k - u \rangle$$
$$= \alpha \langle \nabla f(x_{k+1}), x_{k+1} - z_k \rangle = \frac{(1-\tau)\alpha}{\tau} \langle \nabla f(x_{k+1}), y_k - x_{k+1} \rangle \leq \frac{(1-\tau)\alpha}{\tau} \big( f(y_k) - f(x_{k+1}) \big). \tag{3.2}$$

Above, we used the fact that $\tau(x_{k+1} - z_k) = (1-\tau)(y_k - x_{k+1})$, as well as the convexity of $f(\cdot)$. It is now immediate that by choosing $\frac{1-\tau}{\tau} = \alpha L$ and combining (3.1) and (3.2), we have

**Lemma 3.2** (Coupling). *Letting $\tau \in (0,1)$ satisfy that $\frac{1-\tau}{\tau} = \alpha L$, we have that*

$$\forall u \in Q = \mathbb{R}^n, \quad \alpha \langle \nabla f(x_{k+1}), x_{k+1} - u \rangle \leq \alpha^2 L \big( f(y_k) - f(y_{k+1}) \big) + \big( V_{z_k}(u) - V_{z_{k+1}}(u) \big) \ .$$

It is clear from the above proof that $\tau$ is introduced to precisely balance the objective decrease $f(x_{k+1}) - f(y_{k+1})$, and the (possible) objective increase $f(y_k) - f(x_{k+1})$. This is similar to the "magic number" $K$ discussed in the introduction.

**Finally Convergence Rate.** We telescope inequality Lemma 3.2 for $k = 0, 1, \ldots, T-1$. Setting $\bar{x} \overset{\text{def}}{=} \frac{1}{T} \sum_{k=0}^{T-1} x_k$ and $u = x^*$, we have

$$\alpha T(f(\bar{x}) - f(x^*)) \leq \sum_{k=0}^{T-1} \alpha \langle \partial f(x_k), x_k - x^* \rangle \leq \alpha^2 L \big( f(y_0) - f(y_T) \big) + V_{x_0}(x^*) - V_{x_T}(x^*) \ . \tag{3.3}$$

---

[10] In other words, although a gradient step may decrease the objective from $f(x_{k+1})$ to $f(y_{k+1})$, it may also get the objective increased from $f(y_k)$ to $f(x_{k+1})$.

[11] Indeed, our "thought experiment" in the introduction is conducted *as if* we both had $x_{k+1} = z_k$ and $x_{k+1} = y_k$, and therefore we could arrive at the upcoming (3.3) directly.

Suppose our initial point is of error at most $d$, that is $f(y_0) - f(x^*) \leq d$, and suppose $V_{x_0}(x^*) \leq \Theta$, then (3.3) gives $f(\overline{x}) - f(x^*) \leq \frac{1}{T}(\alpha L d + \Theta/\alpha)$. Choosing $\alpha = \sqrt{\Theta/Ld}$ to be the value that balances the above two terms,[12] we obtain that $f(\overline{x}) - f(x^*) \leq \frac{2\sqrt{L\Theta d}}{T}$. In other words,

$$\text{in } T = 4\sqrt{L\Theta/d} \text{ steps, we can obtain some } \overline{x} \text{ satisfying } f(\overline{x}) - f(x^*) \leq d/2,$$

halving the distance to the optimum. If we restart this entire procedure a few number of times, halving the distance for every run, then we obtain an $\varepsilon$-approximate solution in

$$T = O\big(\sqrt{L\Theta/\varepsilon} + \sqrt{L\Theta/2\varepsilon} + \sqrt{L\Theta/4\varepsilon} + \cdots\big) = O\big(\sqrt{L\Theta/\varepsilon}\big)$$

iterations, matching the same running time of Nesterov's accelerated methods [25–27]. It is important to note here that $\alpha = \sqrt{\Theta/Ld}$ increases as time goes (i.e., as $d$ goes down), and therefore $\tau = \frac{1}{\alpha L + 1}$ decreases as time goes. This lesson instructs us that gradient steps should be given more weights than mirror steps, when it is closer to the optimum.[13]

**Conclusion.** Equipped with the basic knowledge of gradient descent and mirror descent, the above proof is quite straightforward and gives intuition on how the two "magic numbers" $\alpha$ and $\tau$ are selected. However, this simple algorithm has several caveats. First, the value $\alpha$ depends on the knowledge of $\Theta$; second, a good initial distance bound $d$ has to be specified; and third, the algorithm has to be restarted. In the next section, we let $\alpha$ and $\tau$ change gradually across iterations. This overcomes the mentioned caveats, and also extends the above analysis to allow $Q$ to be constrained.

## 4 Final Method with Variable Step Lengths

In this section, we recover the main result of [27] in the constrained case, that is

> **Theorem 4.1.** *If $f(x)$ is $L$-smooth w.r.t. $\|\cdot\|$ on $Q$, and $w(x)$ is 1-strongly convex w.r.t. $\|\cdot\|$ on $Q$, then* `AGM` *outputs $y_T$ satisfying $f(y_T) - f(x^*) \leq 4\Theta L/T^2$, where $\Theta$ is any upper bound on $V_{x_0}(x^*)$.*

We remark here that it is very important to allow the norm $\|\cdot\|$ to be general, rather than focusing on the $\ell_2$-norm as in [26]. See our discussion in Appendix A.1.

Our algorithm `AGM` (see Algorithm 1) starts from $x_0 = y_0 = z_0$. In each step $k = 0, 1, \ldots, T-1$, it computes $x_{k+1} \leftarrow \tau_k z_k + (1 - \tau_k) y_k$ and then
- performs gradient step $y_{k+1} \leftarrow \mathsf{Grad}(x_{k+1})$, and
- performs mirror step $z_{k+1} \leftarrow \mathsf{Mirr}_{z_k}\big(\alpha_{k+1} \nabla f(x_{k+1})\big)$.

Here, $\alpha_{k+1}$ is the step length of mirror descent and will be chosen at the end of this section. The value $\tau_k$ is $\frac{1}{\alpha_{k+1} L}$ which is slightly different from $\frac{1}{\alpha L + 1}$ used in the warm-up case. (This is necessary to capture the constrained case.) Our choice of $\alpha_{k+1}$ will ensure that $\tau_k \in (0, 1]$ for each $k$.

**Convergence Analysis.** We state the analogue of Lemma 3.1 whose proof is in Appendix C:

---

[12] This is essentially the same way to choose $\alpha$ in mirror descent, see (2.3).

[13] One may find this counter-intuitive because when it is closer to the optimum, the observed gradients will become smaller, and therefore mirror steps should perform well due to our conceptual message in the introduction. This understanding is incorrect for two reasons. First, when it is closer to the optimum, the threshold between "large" and "small" gradients also become smaller, so one cannot rely only on mirror steps. Second, when it is closer to the optimum, mirror steps are more 'unstable' and may increase the objective more (in comparison to the current distance to the optimum), and thus should be given less weight.

**Algorithm 1** $\texttt{AGM}(f, w, x_0, T)$

**Input:** $f$ a differentiable and convex function on $Q$ that is $L$-smooth with respect to $\|\cdot\|$;
$w$ the DGF function that is 1-strongly convex with respect to the same $\|\cdot\|$ over $Q$;
$x_0$ some initial point; and $T$ the number of iterations.

**Output:** $y_T$ such that $f(y_T) - f(x^*) \leq \frac{4\Theta L}{T^2}$.

1: $V_x(y) \stackrel{\text{def}}{=} w(y) - \langle \nabla w(x), y - x \rangle - w(x)$.
2: $y_0 \leftarrow x_0, \quad z_0 \leftarrow x_0$.
3: **for** $k \leftarrow 0$ **to** $T - 1$ **do**
4: $\quad \alpha_{k+1} \leftarrow \frac{k+2}{2L}$, and $\tau_k \leftarrow \frac{1}{\alpha_{k+1}L} = \frac{2}{k+2}$.
5: $\quad x_{k+1} \leftarrow \tau_k z_k + (1 - \tau_k) y_k$.
6: $\quad y_{k+1} \leftarrow \mathsf{Grad}(x_{k+1}) \qquad \diamond = \arg\min_{y \in Q}\left\{\frac{L}{2}\|y - x_{k+1}\|^2 + \langle \nabla f(x_{k+1}), y - x_{k+1}\rangle\right\}$
7: $\quad z_{k+1} \leftarrow \mathsf{Mirr}_{z_k}(\alpha_{k+1}\nabla f(x_{k+1})) \quad \diamond = \arg\min_{z \in Q}\left\{V_{z_k}(z) + \langle \alpha_{k+1}\nabla f(x_{k+1}), z - z_k\rangle\right\}$
8: **end for**
9: **return** $y_T$.

**Lemma 4.2.** *If $\tau_k = \frac{1}{\alpha_{k+1}L}$, then it satisfies that for every $u \in Q$,*

$$\alpha_{k+1}\langle \nabla f(x_{k+1}), z_k - u\rangle \leq \alpha_{k+1}^2 L \mathsf{Prog}(x_{k+1}) + V_{z_k}(u) - V_{z_{k+1}}(u)$$
$$\leq \alpha_{k+1}^2 L\big(f(x_{k+1}) - f(y_{k+1})\big) + V_{z_k}(u) - V_{z_{k+1}}(u) \ .$$

We state the analogue of Lemma 3.2, whose proof is slightly different and in Appendix C:

**Lemma 4.3** (Coupling). *For any $u \in Q$,*

$$\big(\alpha_{k+1}^2 L\big) f(y_{k+1}) - \big(\alpha_{k+1}^2 L - \alpha_{k+1}\big) f(y_k) + \big(V_{z_{k+1}}(u) - V_{z_k}(u)\big) \leq \alpha_{k+1} f(u) \ .$$

We are now ready to prove Theorem 4.1:

*Proof of Theorem 4.1.* In order to telescope Lemma 4.3, we only need to set the sequence of $\alpha_k$ so that $\alpha_k^2 L \approx \alpha_{k+1}^2 L - \alpha_{k+1}$ as well as $\tau_k = 1/\alpha_{k+1}L \in (0, 1]$. In our $\texttt{AGM}$, we let $\alpha_k = \frac{k+1}{2L}$ so that $\alpha_k^2 L = \alpha_{k+1}^2 L - \alpha_{k+1} + \frac{1}{4L}$. Summing up Lemma 4.3 for $k = 0, 1, \ldots, T - 1$, we obtain

$$\alpha_T^2 L f(y_T) + \sum_{k=1}^{T-1} \frac{1}{4L} f(y_k) + \big(V_{z_T}(u) - V_{z_0}(u)\big) \leq \sum_{k=1}^{T} \alpha_k f(u) \ .$$

By choosing $u = x^*$, we notice that $\sum_{k=1}^{T} \alpha_k = \frac{T(T+3)}{4L}$, $f(y_k) \geq f(x^*)$, $V_{z_T}(u) \geq 0$ and $V_{z_0}(x^*) \leq \Theta$. Therefore, we obtain

$$\frac{(T+1)^2}{4L^2} L f(y_T) \leq \Big(\frac{T(T+3)}{4L} - \frac{T-1}{4L}\Big) f(x^*) + \Theta \ ,$$

which after simplification implies $f(y_T) \leq f(x^*) + \frac{4\Theta L}{(T+1)^2}$. $\square$

Let us make three remarks.

- $\texttt{AGM}$ is slightly different from [27]: (1) we use Nemirovski's mirror steps instead of dual averaging steps, (2) we allow arbitrary starting points $x_0$, and (3) we use $\tau_k = \frac{2}{k+2}$ rather than $\tau_k = \frac{2}{k+3}$.
- $\texttt{AGM}$ is very different from the perhaps better-known version [26], which is known by some authors as the "momentum method" [32, 41]. Momentum methods do not apply to non-Euclidean settings.

- In Appendix D, we also recover the strong convexity version of accelerated gradient methods [26], and thus linear coupling provides a complete proof of all existing accelerated gradient methods.

## 5 Beyond Accelerated Gradient Methods

Providing an intuitive, yet complete interpretation of accelerated gradient methods is an open question in Optimization [17]. Our result in this paper is one important step towards this general goal. Linear coupling not only gives a reinterpretation of Nesterov's accelerated methods, more importantly, it provides a framework for designing first-order methods in a *bigger agenda*. Since the original version of this paper appeared online, our linear-coupling framework has led to breakthroughs for several problems in computer science. In all such problems, the original Nesterov's accelerated methods do not apply. We illustrate a few examples in this line of research, in order to demonstrate the power and generality of linear coupling.

Recall the key lemmas of gradient and mirror descent in linear coupling (see (3.1)):

$$\text{gradient descent:} \quad f(x_{k+1}) - f(y_{k+1}) \geq \frac{1}{2L}\|\nabla f(x_{k+1})\|_*^2 \tag{5.1}$$

$$\text{mirror descent:} \quad \alpha \langle \nabla f(x_{k+1}), z_k - u \rangle \leq \frac{\alpha^2}{2}\|\nabla f(x_{k+1})\|_*^2 + V_{z_k}(u) - V_{z_{k+1}}(u) \tag{5.2}$$

**Extension 1: Strengthening (5.2) and (5.1).** If $f$ satisfies good properties other than smoothness, one can also develop objective decrease lemma to replace (5.1). In addition, if necessary, a non-strongly convex regularizer can be used in mirror descent to replace (5.2). In either or both such cases, linear coupling can still be used to combine the two methods and obtain faster running times; in contrast, Nesterov's original accelerated methods do not apply.

For example, recent breakthroughs on *positive linear programming (positive LP)* are all based on the above extension of linear coupling [3–5, 22, 42, 43]. For such LPs, the corresponding objective $f$ is intrinsically non-smooth. Some authors including Nesterov himself have applied simple smoothing to turn $f$ into a smooth variant $f'$, and then minimized $f'$ [27]; however, even if Nesterov's accelerated methods are used to minimize $f'$, the resulting running time scales with the problem's *width*, a parameter that can be exponential in input size.[14] In contrast, if linear coupling is used, one can show that $f(x_{k+1}) - f(y_{k+1})$ is lower bounded by a constant times $\sum_j \max\{|\nabla_j f(x_{k+1})|, 1\}^2$ for the original objective $f$ (see [4]). This is a weaker version of (5.1). However, after linear coupling, it leads to a faster algorithm than naively applying Nesterov's accelerated methods on $f'$ in all parameter regimes.

**Extension 2: Three-Point Coupling.** One may naturally consider linearly coupling for more than two vectors. While this is provably unnecessary for minimizing a smooth objective in the full-gradient setting (because accelerated gradient methods are already optimal), it can be very helpful in the *stochastic-gradient* setting.

More specifically, it was a known obstacle in Nesterov's accelerated methods (including our `AGM`) that if the full gradient $\nabla f(x_{k+1})$ is replaced with a random estimator $\widetilde{\nabla}$ whose expectation $\mathbf{E}[\widetilde{\nabla}] = \nabla f(x_{k+1})$, then acceleration disappears in the worst case. Using linear coupling, we can fix this issue by providing the first direct accelerated *stochastic* gradient method. In [1], the author replaced the coupling step $x_{k+1} \leftarrow \tau z_k + (1-\tau)y_k$ with $x_{k+1} \leftarrow \tau_1 z_k + \tau_2 \widetilde{x} + (1-\tau_2-\tau_1)y_k$, where $\widetilde{x}$ is a snapshot point whose full gradient is computed exactly but very infrequently. Such a "three-point" linear coupling provides an accelerated running time because one can combine (5.1), (5.2), together with a so-called variance-reduction inequality [16] all three at once.

---

[14]We recommend interested readers to find detailed discussions in [4] regarding the importance of designing *width-independent* solvers for positive LP. As an illustrative example, in the problem of maximum matching (which can be written as positive LP), the width of the problem is the number of edges in the graph.

**Extension 3: Optimal Sampling Probability.** Nesterov's accelerated methods generalize to coordinate-descent settings, that is, to minimize $f$ that is $L_i$-smooth for each coordinate $i$. The best known coordinate-descent method [21] samples each coordinate $i$ with probability proportional to $L_i$, and is based on a randomized version of Nesterov's original analysis. Using linear coupling, the authors of [6] discovered that one should select $i$ with probability proportional to $\sqrt{L_i}$ for an even faster running time.

To illustrate the reasoning behind this, let us revisit (5.1) and (5.2). In the coordinate-descent setting, if we abbreviate $x_{k+1}$ with $x$, the right hand side of (5.1) simply becomes $\frac{1}{2L_i}(\nabla_i f(x))^2$ if coordinate $i$ is selected. As for (5.2), to ensure its left hand side stays the same in expectation, one should replace $\nabla f(x)$ with $\frac{1}{p_i}\nabla_i f(x)$, where $p_i$ is the probability to select $i$. As a result, the first term on the right hand side of (5.2) becomes $\frac{\alpha^2}{2p_i^2}(\nabla_i f(x))^2$. By comparing these two new terms $\frac{1}{2L_i}(\nabla_i f(x))^2$ and $\frac{\alpha^2}{2p_i^2}(\nabla_i f(x))^2$, we immediately notice that $p_i$ had better be proportional to $\sqrt{L_i}$ in order for the two terms to cancel. This simple idea, fully motivated from linear coupling, leads to the fastest accelerated coordinate-descent method [6].

**Extension 4: Supporting Non-Convexity.** Consider objectives $f$ that are not even convex but still smooth. For instance, neural network training objectives fall into this class if smoothed activation functions are used. In such a case, both (5.1) and (5.2) remain true. However, when coupling the two steps, we cannot claim $\langle \nabla f(x_{k+1}), x_{k+1} - u \rangle \geq f(x_{k+1}) - f(u)$ because there is no convexity. In [2], the authors discovered that one can use the quadratic lower bound $\langle \nabla f(x_{k+1}), x_{k+1} - u \rangle \geq f(x_{k+1}) - f(u) - \frac{L}{2}\|x_{k+1} - u\|^2$ to replace convexity arguments, and still perform a weaker version of linear coupling. This leads to a stochastic algorithm that converges to approximate saddle-points,[15] outperforming both gradient descent and stochastic gradient descent, the only two known first-order methods with provably convergence guarantees.

## Acknowledgements

We thank Jon Kelner and Yin Tat Lee for helpful conversations, and Aaditya Ramdas for pointing out a typo in the previous version of this paper.

This material is based upon work partly supported by the National Science Foundation under Grant CCF-1319460 and by a Simons Graduate Student Award under grant no. 284059.

---

[15]Recall that in general non-convex optimization one can only hope for converging to saddle-points.

# Appendix

## A Several Remarks on First-Order Methods

### A.1 Importance of Non-Euclidean Norms

Let us use a simple example to illustrate the importance of allowing arbitrary norms in studying first-order methods.

Consider the saddle point problem of $\min_{x \in \Delta_n} \max_{y \in \Delta_m} y^T A x$, where $A$ is an $m \times n$ matrix, $\Delta_n = \{x \in \mathbb{R}^n : x \geq 0 \wedge \mathbb{1}^T x = 1\}$ is the unit simplex in $\mathbb{R}^n$, and $\Delta_m = \{y \in \mathbb{R}^m : y \geq 0 \wedge \mathbb{1}^T y = 1\}$. This problem is important to study because it captures packing and covering linear programs that have wide applications in many areas of computer science (see the survey of [8]).

Letting $\mu = \frac{\varepsilon}{2 \log m}$, Nesterov has shown that the following objective

$$f_\mu(x) \stackrel{\text{def}}{=} \mu \log \Big( \frac{1}{m} \sum_{j=1}^m \exp^{\frac{1}{\mu}(Ax)_j} \Big) \ ,$$

when optimized over $x \in \Delta_n$, can yield an additive $\varepsilon/2$ solution to the original saddle point problem [27].

This $f_\mu(x)$ is proven to be $\frac{1}{\mu}$-smooth with respect to the $\ell_1$-norm over $\Delta_n$, if all the entries of $A$ are between $[-1, 1]$. Instead, $f_\mu(x)$ is $\frac{1}{\mu}$-smooth with respect to the $\ell_2$-norm over $\Delta_n$, *only if* the sum of squares of every row of $A$ is at most 1. This $\ell_2$ condition is certainly stronger and less natural than the $\ell_1$ condition, and the $\ell_1$ condition one leads to the fastest (approximate) width-dependent positive LP solver [27].

Different norm conditions also yield different gradient and mirror descent steps. For instance, in the $\ell_1$-norm case, the gradient step is $x' \leftarrow \arg\min_{x' \in \Delta_n} \big\{ \frac{1}{2} \|x' - x\|_1^2 + \alpha \langle \nabla f_\mu(x), x' - x \rangle \big\}$, and the mirror step is $x' \leftarrow \arg\min_{x' \in \Delta_n} \big\{ \sum_{i \in [n]} x'_i \log \frac{x'_i}{x_i} + \alpha \langle \nabla f_\mu(x), x' - x \rangle \big\}$. In the $\ell_2$-norm case, gradient and mirror steps are both of the form $x' \leftarrow \arg\min_{x' \in \Delta_n} \big\{ \frac{1}{2} \|x' - x\|_2^2 + \alpha \langle \nabla f_\mu(x), x' - x \rangle \big\}$.

As another example, [35] has shown that the $\ell_1$ norm, instead of the $\ell_2$ one, is crucial when computing the minimum enclosing ball of points. One can find other applications as well in [27] for the use of non-Euclidean norms, and an interesting example of $\ell_\infty$-norm gradient descent for nearly-linear time maximum flow in [18].

It is now important to note that, the methods in [25, 26] work only for the $\ell_2$-norm case, and it is not clear how the proof can be generalized to other norms until [27]. Some other proofs (such as [13]) only work for the $\ell_2$-norm because the mirror steps are described as (a scaled version of) gradient steps.

### A.2 Folklore Relationship Between Multiplicative Weight Updates and Mirror Descent

The multiplicative weight update (MWU) method (see the survey of Arora, Hazan and Kale [8]) is a simple method that has been repeatedly discovered in theory of computation, machine learning, optimization, and game theory. The setting of this method is the following.

Let $\Delta_n = \{x \in \mathbb{R}^n : x \geq 0 \wedge \mathbb{1}^T x = 1\}$ be the unit simplex in $\mathbb{R}^n$, and we call any vector in $\Delta_n$ an *action*. A player is going to play $T$ actions $x_0, \ldots, x_{T-1} \in \Delta_n$ in a row; only after playing $x_k$, the player observes a loss vector $\ell_k \in \mathbb{R}^n$ that may depend on $x_k$, and suffers from a loss value $\langle \ell_k, x_k \rangle$. The MWU method ensures that, if $\|\ell_k\|_\infty \leq \rho$ for all $k \in [T]$, then the player has an (adaptive) strategy to choose the actions such that the average *regret* is bounded:

$$\frac{1}{T}\Big(\sum_{i=0}^{T-1}\langle \ell_k, x_k\rangle - \min_{u\in\Delta_n}\sum_{i=0}^{T-1}\langle \ell_k, u\rangle\Big) \leq O\Big(\frac{\rho\sqrt{\log n}}{\sqrt{T}}\Big) \ . \quad \text{(A.1)}$$

The left hand side is called the average regret because it is the (average) difference between the suffered loss $\sum_{i=0}^{T-1}\langle \ell_k, x_k\rangle$, and the loss $\sum_{i=0}^{T-1}\langle \ell_k, u\rangle$ of the best action $u \in \Delta_n$ in hindsight. Another way to interpret (A.1) is to state that we can obtain an average regret of $\varepsilon$ using $T = O(\frac{\rho^2 \log n}{\varepsilon^2})$ rounds.

The above result can be proven directly using mirror descent. Letting $w(x) \stackrel{\text{def}}{=} \sum_i x_i \log x_i$ be the entropy DGF over the simplex $Q = \Delta_n$, and its corresponding Bregman divergence $V_x(x') \stackrel{\text{def}}{=} \sum_{i\in[n]} x'_i \log \frac{x'_i}{x_i}$, we consider the following update rule.

Start from $x_0 = (1/n, \ldots, 1/n)$, and update $x_{k+1} = \mathsf{Mirr}_{x_k}(\alpha\ell_k)$, or equivalently, $x_{k+1,i} = x_{k,i} \cdot \exp^{-\alpha\ell_{k,i}}/Z_k$, where $Z_k > 0$ is the normalization factor that equals to $\sum_{i=1}^n x_{k,i} \cdot \exp^{-\alpha\ell_{k,i}}$.[16] Then, the mirror-descent guarantee (2.2) implies that[17]

$$\forall u \in \Delta_n, \quad \alpha\langle \ell_k, x_k - u\rangle \leq \frac{\alpha^2}{2}\|\ell_k\|_\infty^2 + V_{x_k}(u) - V_{x_{k+1}}(u) \ .$$

After telescoping the above inequality for all $k = 0, 1, \ldots, T-1$, and using the upper bounds $\|\ell(x_k)\|_\infty \leq \rho$ and $V_{x_0}(u) \leq \log n$, we obtain that for all $u \in \Delta_n$,

$$\frac{1}{T}\sum_{k=0}^{T-1}\langle \ell_k, x_k - u\rangle \leq \frac{\alpha\rho^2}{2} + \frac{\log n}{\alpha T} \ .$$

Setting $\alpha = \frac{\sqrt{\log n}}{\rho\sqrt{T}}$ we arrive at the desired average regret bound (A.1).

In sum, we have re-deduced the MWU method from mirror descent, and the above proof is quite different from most of the classical analysis of MWU (e.g., [7, 8, 14, 34]). It can be generalized to solve the matrix version of MWU [8, 33], as well as to incorporate the width-reduction technique [8, 34]. We ignore such extensions here because they are outside the scope of this paper.

## A.3 Deducing the Mirror-Descent Guarantee via Gradient Descent

In this section, we re-deduce the convergence rate of mirror descent from gradient descent. In particular, we show that the dual averaging steps are equivalent to gradient steps on the Fenchel dual of the regularized regret, and deduce the same convergence bound as (2.4). (Similar proof can also be obtained for mirror steps but is notationally more involved.)

Given a sequence of points $x_0, \ldots, x_{T-1} \in Q$, the (scaled) regret with respect to any point $u \in Q$ is $R(x_0, \ldots, x_{T-1}, u) \stackrel{\text{def}}{=} \sum_{i=0}^{T-1}\alpha\langle \partial f(x_i), x_i - u\rangle$. Since it satisfies that $\alpha T \cdot (f(\overline{x}) - f(u)) \leq$ $R(x_0, \ldots, x_{T-1}, u)$, the average regret (after scaling) upper bounds on the distance between any point $f(u)$ and the average $\overline{x} = \frac{1}{T}(x_0 + \cdots + x_{T-1})$. Consider now the regularized regret

$$\widehat{R}(x_0, \ldots, x_{T-1}) \stackrel{\text{def}}{=} \max_{u \in Q} \Big\{ \sum_{i=0}^{T-1} \alpha \langle \partial f(x_i), x_i - u \rangle - w(u) \Big\} \,,$$

and we can rewrite it using the Fenchel dual $w^*(\lambda) \stackrel{\text{def}}{=} \max_{u \in Q} \{\langle \lambda, u \rangle - w(u)\}$ of $w(\cdot)$:

$$\widehat{R}(x_0, \ldots, x_{T-1}) = w^*\Big( -\alpha \sum_{i=0}^{T-1} \partial f(x_i) \Big) + \sum_{i=0}^{T-1} \alpha \langle \partial f(x_i), x_i \rangle \,.$$

---

[16]This version of the MWU is often known as the Hedge rule [14]. Another commonly used version is to choose $x_{k+1,i} = \frac{x_{k,i}(1-\alpha\ell_{k,i})}{Z_k}$. Since $e^{-t} \approx 1 - t$ whenever $|t|$ is small and our choice of $\alpha$ will make sure that $|\alpha\ell_{k,i}| \ll 1$, this is essentially identical to the Hedge rule.

[17]To be precise, we have replaced $\partial f(x_k)$ with $\ell_k$. It is easy to see from the proof of (2.2) that this loss vector $\ell_k$ does not need to come from the subgradient of some objective $f(\cdot)$.

The classical theory of Fenchel duality tells us that $w^*(\lambda)$ is 1-smooth with respect to the dual norm $\|\cdot\|_*$, because $w(\cdot)$ is 1-strongly convex with respect to $\|\cdot\|$. We also have $\nabla w^*(\lambda) = \arg\max_{u \in Q}\{\langle \lambda, u \rangle - w(u)\}$. (See for instance [36].)

With enough notations introduced, let us now minimize $\widehat{R}$ by intelligently selecting $x_0, \ldots, x_{T-1}$. Perhaps a little counter-intuitively, we start from $x_0 = \cdots = x_{T-1} = x^*$ and accordingly $\partial f(x^*) = 0$ (if there are multiple subgradients at $x^*$, choose the zero one). This corresponds to a regret value of zero and a regularized regret $\widehat{R}(x^*, \ldots, x^*) = w^*(0) = -\min_{u \in Q}\{w(u)\}$.

Next, we choose the values of $x_0, \ldots, x_{T-1}$ one by one. We choose $x_0 = \arg\min_{u \in Q}\{w(u)\}$ as the starting point.[18] Suppose that the values of $x_0, \ldots, x_{k-1}$ are already determined, and we are ready to pick $x_k \in Q$. Let us compute the changes in the regularized regret as a function of $x_k$:

$$\Delta \widehat{R} = \widehat{R}(x_0, \ldots, x_k, x^*, \ldots, x^*) - \widehat{R}(x_0, \ldots, x_{k-1}, x^*, \ldots, x^*)$$

$$= w^*\Big( -\alpha \sum_{i=0}^{k} \partial f(x_i) \Big) - w^*\Big( -\alpha \sum_{i=0}^{k-1} \partial f(x_i) \Big) + \alpha \langle \partial f(x_k), x_k \rangle$$

$$\leq \Big\langle \nabla w^*\Big( -\alpha \sum_{i=0}^{k-1} \partial f(x_i) \Big), -\alpha \partial f(x_k) \Big\rangle + \frac{1}{2}\|\alpha \partial f(x_k)\|_*^2 + \alpha \langle \partial f(x_k), x_k \rangle \,. \quad \text{(A.2)}$$

Here, the last inequality is because $w^*(a) - w^*(b) \leq \langle \nabla w^*(b), a - b \rangle + \frac{1}{2}\|a - b\|_*^2$, owing to the smoothness of $w^*(\cdot)$. At this moment, it is clear to see that if one chooses

$$x_k = \nabla w^*\Big( -\alpha \sum_{i=0}^{k-1} \partial f(x_i) \Big) = \arg\min_{u \in Q} \Big\{ w(u) + \sum_{i=0}^{k-1} \alpha \langle \partial f(x_i), u \rangle \Big\} \,,$$

the first and third terms in (A.2) cancel out, and we obtain $\Delta \widehat{R} \leq \frac{1}{2}\|\alpha \partial f(x_k)\|_*^2$. In other words, the regularized regret increases by no more than $\frac{1}{2}\|\alpha \partial f(x_k)\|_*^2 \leq \alpha^2 \rho^2 / 2$ in each step, so in the end we have $\widehat{R}(x_0, \ldots, x_{T-1}) \leq -w(x_0) + \alpha^2 \rho^2 T / 2$.

In sum, by the definition of the regularized regret, we have

$$\alpha T \cdot (f(\overline{x}) - f(x^*)) - w(x^*) \leq \sum_{i=0}^{T-1} \alpha \langle \partial f(x_i), x_i - x^* \rangle - w(x^*) \leq \widehat{R}(x_0, \ldots, x_{T-1}) \leq -w(x_0) + \frac{\alpha^2 \rho^2 T}{2} \,.$$

This implies the following upper bound on the optimality of $f(\overline{x})$

$$f(\overline{x}) - f(x^*) \leq \frac{\alpha \rho^2}{2} + \frac{w(x^*) - w(x_0)}{\alpha T} = \frac{\alpha \rho^2}{2} + \frac{V_{x_0}(x^*)}{\alpha T} \leq \frac{\alpha \rho^2}{2} + \frac{\Theta}{\alpha T} \,.$$

---

[18]Dual averaging steps typically demand the first point $x_0$ to be at the minimum of the regularizer $w(\cdot)$, because that leads to the cleanest analysis. This can be relaxed to allow an arbitrary starting point.

Finally, choosing $\alpha = \frac{\sqrt{2\Theta}}{\rho \cdot \sqrt{T}}$ to be the step length, we arrive at $f(\bar{x}) - f(x^*) \leq \frac{\sqrt{2\Theta} \cdot \rho}{\sqrt{T}}$, which is the same convergence rate as (2.4).

## B  Missing Proof of Section 2

For the sake of completeness, we provide self-contained proofs of the mirror descent and mirror descent guarantees in this section.

### B.1  Missing Proof for Gradient Descent

> **Gradient Descent Guarantee**
>
> $$f(\mathsf{Grad}(x)) \leq f(x) - \mathsf{Prog}(x) \qquad (2.1)$$
>
> or in the special case when $Q = \mathbb{R}^n$ $\qquad f(\mathsf{Grad}(x)) \leq f(x) - \frac{1}{2L}\|\nabla f(x)\|_*^2 \;.$

*Proof.* [19] Letting $\widetilde{x} = \mathsf{Grad}(x)$, we prove the first inequality by

$$\mathsf{Prog}(x) = -\min_{y \in Q}\left\{\frac{L}{2}\|y - x\|^2 + \langle \nabla f(x), y - x\rangle\right\} = -\left(\frac{L}{2}\|\widetilde{x} - x\|^2 + \langle \nabla f(x), \widetilde{x} - x\rangle\right)$$

$$= f(x) - \left(\frac{L}{2}\|\widetilde{x} - x\|^2 + \langle \nabla f(x), \widetilde{x} - x\rangle + f(x)\right) \leq f(x) - f(\widetilde{x}) \;.$$

Here, the last inequality is a consequence of the smoothness assumption: for any $x, y \in Q$,

$$f(y) - f(x) = \int_{\tau=0}^{1} \langle \nabla f(x + \tau(y - x)), y - x\rangle d\tau$$

$$= \langle \nabla f(x), y - x\rangle + \int_{\tau=0}^{1} \langle \nabla f(x + \tau(y - x)) - \nabla f(x), y - x\rangle d\tau$$

$$\leq \langle \nabla f(x), y - x\rangle + \int_{\tau=0}^{1} \|\nabla f(x + \tau(y - x)) - \nabla f(x)\|_* \cdot \|y - x\| d\tau$$

$$\leq \langle \nabla f(x), y - x\rangle + \int_{\tau=0}^{1} \tau L \|y - x\| \cdot \|y - x\| d\tau = \langle \nabla f(x), y - x\rangle + \frac{L}{2}\|y - x\|^2$$

The second inequality follows because in the special case of $Q = \mathbb{R}^n$, we have

$$\mathsf{Prog}(x) = -\min_{y \in Q}\left\{\frac{L}{2}\|y - x\|^2 + \langle \nabla f(x), y - x\rangle\right\} = \frac{1}{2L}\|\nabla f(x)\|_*^2 \;. \qquad \square$$

**Fact B.1** (Gradient Descent Convergence). *Let $f(x)$ be a convex, differentiable function that is $L$-smooth with respect to $\|\cdot\|$ on $Q = \mathbb{R}^n$, and $x_0$ any initial point in $Q$. Consider the sequence of $T$ gradient steps $x_{k+1} \leftarrow \mathsf{Grad}(x_k)$, then the last point $x_T$ satisfies that*

$$f(x_T) - f(x^*) \leq O\left(\frac{LR^2}{T}\right) \;,$$

*where $R = \max_{x: f(x) \leq f(x_0)} \|x - x^*\|$, and $x^*$ is any minimizer of $f$.*

---
[19]This proof can be found for instance in the textbook [26].

*Proof.* [20] Recall that we have $f(x_{k+1}) \leq f(x_k) - \frac{1}{2L}\|\nabla f(x_k)\|_*^2$ from (2.1). Furthermore, by the convexity of $f$ and Cauchy-Schwarz we have

$$f(x_k) - f(x^*) \leq \langle \nabla f(x_k), x_k - x^* \rangle \leq \|\nabla f(x_k)\|_* \cdot \|x_k - x^*\| \leq R \cdot \|\nabla f(x_k)\|_* .$$

Letting $D_k = f(x_k) - f(x^*)$ denote the distance to the optimum at iteration $k$, we now obtain two relationships $D_k - D_{k+1} \geq \frac{1}{2L}\|\nabla f(x_k)\|_*^2$ as well as $D_k \leq R \cdot \|\nabla f(x_k)\|_*$. Combining these two, we get

$$D_k^2 \leq 2LR^2(D_k - D_{k+1}) \implies \frac{D_k}{D_{k+1}} \leq 2LR^2\Big(\frac{1}{D_{k+1}} - \frac{1}{D_k}\Big) .$$

Noticing that $D_k \geq D_{k+1}$ because our objective only decreases at every round, we obtain that $\frac{1}{D_{k+1}} - \frac{1}{D_k} \geq \frac{1}{2LR^2}$. Finally, we conclude that at round $T$, we must have $\frac{1}{D_T} \geq \frac{T}{2LR^2}$, finishing the proof that $f(x_T) - f(x^*) \leq \frac{2LR^2}{T}$. □

## B.2 Missing Proof for Mirror Descent

> **Mirror Descent Guarantee**
>
> If $x_{k+1} = \mathsf{Mirr}_{x_k}(\alpha \cdot \partial f(x_k))$, then
>
> $$\forall u \in Q, \quad \alpha(f(x_k) - f(u)) \leq \alpha\langle \partial f(x_k), x_k - u\rangle \leq \frac{\alpha^2}{2}\|\partial f(x_k)\|_*^2 + V_{x_k}(u) - V_{x_{k+1}}(u) . \quad (2.2)$$

*Proof.* [21] we compute that

$$\alpha\langle \partial f(x_k), x_k - u\rangle = \langle \alpha \partial f(x_k), x_k - x_{k+1}\rangle + \langle \alpha \partial f(x_k), x_{k+1} - u\rangle$$

$$\stackrel{①}{\leq} \langle \alpha \partial f(x_k), x_k - x_{k+1}\rangle + \langle -\nabla V_{x_k}(x_{k+1}), x_{k+1} - u\rangle$$

$$\stackrel{②}{=} \langle \alpha \partial f(x_k), x_k - x_{k+1}\rangle + V_{x_k}(u) - V_{x_{k+1}}(u) - V_{x_k}(x_{k+1})$$

$$\stackrel{③}{\leq} \Big(\langle \alpha \partial f(x_k), x_k - x_{k+1}\rangle - \frac{1}{2}\|x_k - x_{k+1}\|^2\Big) + \big(V_{x_k}(u) - V_{x_{k+1}}(u)\big)$$

$$\stackrel{④}{\leq} \frac{\alpha^2}{2}\|\partial f(x_k)\|_*^2 + \big(V_{x_k}(u) - V_{x_{k+1}}(u)\big)$$

Here, ① is due to the minimality of $x_{k+1} = \arg\min_{x \in Q}\{V_{x_k}(x) + \langle \alpha \partial f(x_k), x\rangle\}$, which implies that $\langle \nabla V_{x_k}(x_{k+1}) + \alpha \partial f(x_k), u - x_{k+1}\rangle \geq 0$ for all $u \in Q$. ② is due to the triangle equality of Bregman divergence.[22] ③ is because $V_x(y) \geq \frac{1}{2}\|x - y\|^2$ by the strong convexity of the DGF $w(\cdot)$. ④ is by Cauchy-Schwarz. □

---

[20] Our proof follows almost directly from [26], but he only uses the Euclidean $\ell_2$ norm.
[21] This proof can be found for instance in the textbook [9].
[22] That is,

$$\forall x, y \geq 0, \quad \langle -\nabla V_x(y), y - u\rangle = \langle \nabla w(x) - \nabla w(y), y - u\rangle$$
$$= (w(u) - w(x) - \langle \nabla w(x), u - x\rangle) - (w(u) - w(y) - \langle \nabla w(y), u - y\rangle)$$
$$\quad - (w(y) - w(x) - \langle \nabla w(x), y - x\rangle)$$
$$= V_x(u) - V_y(u) - V_x(y) .$$

# C Missing Proofs of Section 4

**Lemma 4.2.** *If $\tau_k = \frac{1}{\alpha_{k+1}L}$, then it satisfies that for every $u \in Q$,*

$$\alpha_{k+1}\langle \nabla f(x_{k+1}), z_k - u\rangle \overset{①}{\leq} \alpha_{k+1}^2 L \mathsf{Prog}(x_{k+1}) + V_{z_k}(u) - V_{z_{k+1}}(u)$$

$$\overset{②}{\leq} \alpha_{k+1}^2 L\big(f(x_{k+1}) - f(y_{k+1})\big) + V_{z_k}(u) - V_{z_{k+1}}(u) \ .$$

*Proof.* The second inequality ② is again from the gradient descent guarantee $f(x_{k+1}) - f(y_{k+1}) \geq \mathsf{Prog}(x_{k+1})$. To prove ①, we first write down the key inequality of mirror-descent analysis (whose proof is identical to that of (2.2))

$$\alpha_{k+1}\langle \nabla f(x_{k+1}), z_k - u\rangle = \langle \alpha_{k+1}\nabla f(x_{k+1}), z_k - z_{k+1}\rangle + \langle \alpha_{k+1}\nabla f(x_{k+1}), z_{k+1} - u\rangle$$

$$\overset{①}{\leq} \langle \alpha_{k+1}\nabla f(x_{k+1}), z_k - z_{k+1}\rangle + \langle -\nabla V_{z_k}(z_{k+1}), z_{k+1} - u\rangle$$

$$\overset{②}{=} \langle \alpha_{k+1}\nabla f(x_{k+1}), z_k - z_{k+1}\rangle + V_{z_k}(u) - V_{z_{k+1}}(u) - V_{z_k}(z_{k+1})$$

$$\overset{③}{\leq} \Big(\langle \alpha_{k+1}\nabla f(x_{k+1}), z_k - z_{k+1}\rangle - \frac{1}{2}\|z_k - z_{k+1}\|^2\Big) + \big(V_{z_k}(u) - V_{z_{k+1}}(u)\big)$$

Here, ① is due to the minimality of $z_{k+1} = \arg\min_{z \in Q}\{V_{z_k}(z) + \langle \alpha_{k+1}\nabla f(x_{k+1}), z\rangle\}$, which implies that $\langle \nabla V_{z_k}(z_{k+1}) + \alpha_{k+1}\nabla f(x_{k+1}), u - z_{k+1}\rangle \geq 0$ for all $u \in Q$. ② is due to the triangle equality of Bregman divergence (see Footnote 22 in Appendix B). ③ is because $V_x(y) \geq \frac{1}{2}\|x - y\|^2$ by the strong convexity of the $w(\cdot)$.

If one stops here and uses Cauchy-Shwartz $\langle \alpha_{k+1}\nabla f(x_{k+1}), z_k - z_{k+1}\rangle - \frac{1}{2}\|z_k - z_{k+1}\|^2 \leq \frac{\alpha_{k+1}^2}{2}\|\nabla f(x_{k+1})\|_*^2$, he will get the desired inequality in the special case of $Q = \mathbb{R}^n$, because $\mathsf{Prog}(x_{k+1}) = \frac{1}{2L}\|\nabla f(x_{k+1})\|_*^2$ from (2.1).

For the general unconstrained case, we need to use the special choice of $\tau_k = 1/\alpha_{k+1}L$ follows. Letting $v \overset{\text{def}}{=} \tau_k z_{k+1} + (1 - \tau_k)y_k \in Q$ so that $x_{k+1} - v = (\tau_k z_k + (1 - \tau_k)y_k) - v = \tau_k(z_k - z_{k+1})$, we have

$$\langle \alpha_{k+1}\nabla f(x_{k+1}), z_k - z_{k+1}\rangle - \frac{1}{2}\|z_k - z_{k+1}\|^2$$

$$= \langle \frac{\alpha_{k+1}}{\tau_k}\nabla f(x_{k+1}), x_{k+1} - v\rangle - \frac{1}{2\tau_k^2}\|x_{k+1} - v\|^2$$

$$= \alpha_{k+1}^2 L\left(\langle \nabla f(x_{k+1}), x_{k+1} - v\rangle - \frac{L}{2}\|x_{k+1} - v\|^2\right) \leq \alpha_{k+1}^2 L \mathsf{Prog}(x_{k+1})$$

where the last inequality is from the definition of $\mathsf{Prog}(x_{k+1})$. $\square$

**Lemma 4.3** (Coupling). *For any $u \in Q$,*

$$\big(\alpha_{k+1}^2 L\big)f(y_{k+1}) - \big(\alpha_{k+1}^2 L - \alpha_{k+1}\big)f(y_k) + \big(V_{z_{k+1}}(u) - V_{z_k}(u)\big) \leq \alpha_{k+1}f(u) \ .$$

*Proof.* We deduce the following sequence of inequalities

$$\begin{aligned}
&\alpha_{k+1}\big(f(x_{k+1}) - f(u)\big) \\
&\leq \alpha_{k+1}\langle \nabla f(x_{k+1}), x_{k+1} - u\rangle \\
&= \alpha_{k+1}\langle \nabla f(x_{k+1}), x_{k+1} - z_k\rangle + \alpha_{k+1}\langle \nabla f(x_{k+1}), z_k - u\rangle \\
&\stackrel{①}{=} \frac{(1-\tau_k)\alpha_{k+1}}{\tau_k}\langle \nabla f(x_{k+1}), y_k - x_{k+1}\rangle + \alpha_{k+1}\langle \nabla f(x_{k+1}), z_k - u\rangle \\
&\stackrel{②}{\leq} \frac{(1-\tau_k)\alpha_{k+1}}{\tau_k}\big(f(y_k) - f(x_{k+1})\big) + \alpha_{k+1}\langle \nabla f(x_{k+1}), z_k - u\rangle \\
&\stackrel{③}{\leq} \frac{(1-\tau_k)\alpha_{k+1}}{\tau_k}\big(f(y_k) - f(x_{k+1})\big) + \alpha_{k+1}^2 L\big(f(x_{k+1}) - f(y_{k+1})\big) + V_{z_k}(u) - V_{z_{k+1}}(u) \\
&\stackrel{④}{=} \big(\alpha_{k+1}^2 L - \alpha_{k+1}\big)f(y_k) - \big(\alpha_{k+1}^2 L\big)f(y_{k+1}) + \alpha_{k+1}f(x_{k+1}) + \big(V_{z_k}(u) - V_{z_{k+1}}(u)\big)
\end{aligned}$$

Here, ① uses the choice of $x_{k+1}$ that satisfies $\tau_k(x_{k+1} - z_k) = (1-\tau_k)(y_k - x_{k+1})$; ② is by the convexity of $f(\cdot)$ and $1 - \tau_k \geq 0$; ③ uses Lemma 4.2; and ④ uses the choice of $\tau_k = 1/\alpha_{k+1}L$. □

# D  Strong Convexity Version of Accelerated Gradient Method

When the objective $f(\cdot)$ is both $\sigma$-strongly convex and $L$-smooth with respect to the same norm $\|\cdot\|_2$, another version of accelerated gradient method exists and achieves a $\log(1/\varepsilon)$ convergence rate [26]. We show in this section that, our method $\mathtt{AGM}(f, w, x_0, T)$ can be used to recover that strong-convexity accelerated method in one of the two ways. Therefore, the gradient-mirror coupling interpretation behind our paper still applies to the strong-convexity accelerated method.

One way to recover the strong-convexity accelerated method is to replace the use of the mirror-descent analysis on the regret term by its strong-convexity counterpart (also known as logarithmic-regret analysis, see for instance [15, 37]). This would incur some different parameter choices on $\alpha_k$ and $\tau_k$, and results in an algorithm similar to that of [26].

Another, but simpler way is to recursively apply Theorem 4.1. In light of the definition of strong convexity and Theorem 4.1, we have

$$\frac{\sigma}{2}\|y_T - x^*\|_2^2 \leq f(y_T) - f(x^*) \leq \frac{4 \cdot \frac{1}{2}\|x_0 - x^*\|_2^2 \cdot L}{T^2} \ .$$

In particular, in every $T = T_0 \stackrel{\text{def}}{=} \sqrt{8L/\sigma}$ iterations, we can halve the distance $\|y_T - x^*\|_2^2 \leq \frac{1}{2}\|x_0 - x^*\|_2^2$. If we repeatedly invoke $\mathtt{AGM}(f, w, \cdot, T_0)$ a sequence of $\ell$ times, each time feeding the initial vector $x_0$ with the previous output $y_{T_0}$, then in the last run of the $T_0$ iterations, we have

$$f(y_{T_0}) - f(x^*) \leq \frac{4 \cdot \frac{1}{2^\ell}\|x_0 - x^*\|_2^2 \cdot L}{T_0^2} = \frac{1}{2^{\ell+1}}\|x_0 - x^*\|_2^2 \cdot \sigma \ .$$

By choosing $\ell = \log\big(\frac{\|x_0 - x^*\|_2^2 \cdot \sigma}{\varepsilon}\big)$, we conclude that

> **Corollary D.1.** *If $f(\cdot)$ is both $\sigma$-strongly convex and $L$-smooth with respect to $\|\cdot\|_2$, in a total of $T = O\big(\sqrt{\frac{L}{\sigma}} \cdot \log\big(\frac{\|x_0 - x^*\|_2^2 \cdot \sigma}{\varepsilon}\big)\big)$ iterations, we can obtain some $x$ such that $f(x) - f(x^*) \leq \varepsilon$.*

This is slightly better than the result $O\big(\sqrt{\frac{L}{\sigma}} \cdot \log\big(\frac{\|x_0 - x^*\|_2^2 \cdot L}{\varepsilon}\big)\big)$ in Theorem 2.2.2 of [26].

We remark here that O'Donoghue and Candès [32] have studied some heuristic adaptive restarting techniques which suggest that the above (and other) restarting version of the accelerated method practically outperforms the original method of Nesterov.

# References

[1] Zeyuan Allen-Zhu. Katyusha: Accelerated Variance Reduction for Faster SGD. *ArXiv e-prints*, abs/1603.05953, March 2016.

[2] Zeyuan Allen-Zhu and Elad Hazan. Variance Reduction for Faster Non-Convex Optimization. In *ICML*, 2016.

[3] Zeyuan Allen-Zhu, Yin Tat Lee, and Lorenzo Orecchia. Using optimization to obtain a width-independent, parallel, simpler, and faster positive SDP solver. In *SODA*, 2016.

[4] Zeyuan Allen-Zhu and Lorenzo Orecchia. Nearly-Linear Time Positive LP Solver with Faster Convergence Rate. In *STOC*, 2015.

[5] Zeyuan Allen-Zhu and Lorenzo Orecchia. Using optimization to break the epsilon barrier: A faster and simpler width-independent algorithm for solving positive linear programs in parallel. In *SODA*, 2015.

[6] Zeyuan Allen-Zhu, Peter Richtárik, Zheng Qu, and Yang Yuan. Even faster accelerated coordinate descent using non-uniform sampling. In *ICML*, 2016.

[7] Sanjeev Arora, Elad Hazan, and Satyen Kale. Fast Algorithms for Approximate Semidefinite Programming using the Multiplicative Weights Update Method. In *FOCS*, pages 339–348. IEEE, 2005.

[8] Sanjeev Arora, Elad Hazan, and Satyen Kale. The Multiplicative Weights Update Method: a Meta-Algorithm and Applications. *Theory of Computing*, 8:121–164, 2012.

[9] Aharon Ben-Tal and Arkadi Nemirovski. *Lectures on Modern Convex Optimization*. Society for Industrial and Applied Mathematics, January 2013.

[10] Sébastien Bubeck, Yin Tat Lee, and Mohit Singh. A geometric alternative to Nesterov's accelerated gradient descent. *ArXiv e-prints*, abs/1506.08187, June 2015.

[11] Ofer Dekel, Ran Gilad-Bachrach, Ohad Shamir, and Lin Xiao. Optimal distributed online prediction using mini-batches. *The Journal of Machine Learning Research*, 13(1):165–202, 2012.

[12] John Duchi, Shai Shalev-Shwartz, Yoram Singer, and Ambuj Tewari. Composite Objective Mirror Descent. In *COLT*, 2010.

[13] Olivier Fercoq and Peter Richtárik. Accelerated, parallel, and proximal coordinate descent. *SIAM Journal on Optimization*, 25(4):1997–2023, 2015. First appeared on ArXiv 1312.5799 in 2013.

[14] Yoav Freund and Robert E Schapire. A desicion-theoretic generalization of on-line learning and an application to boosting. In *Computational learning theory*, pages 23–37. Springer, 1995.

[15] Elad Hazan, Amit Agarwal, and Satyen Kale. Logarithmic regret algorithms for online convex optimization. *Machine Learning*, 69(2-3):169–192, August 2007.

[16] Rie Johnson and Tong Zhang. Accelerating stochastic gradient descent using predictive variance reduction. In *Advances in Neural Information Processing Systems*, NIPS 2013, pages 315–323, 2013.

[17] Anatoli Juditsky. Convex optimization ii: Algorithms. Lecture notes, November 2013.

[18] Jonathan A. Kelner, Yin Tat Lee, Lorenzo Orecchia, and Aaron Sidford. An Almost-Linear-Time Algorithm for Approximate Max Flow in Undirected Graphs, and its Multicommodity Generalizations. In *SODA*, April 2014.

[19] Guanghui Lan. An optimal method for stochastic composite optimization. *Mathematical Programming*, 133(1-2):365–397, January 2011.

[20] Yin Tat Lee, Satish Rao, and Nikhil Srivastava. A new approach to computing maximum flows using electrical flows. In *STOC*, page 755, New York, New York, USA, 2013.

[21] Yin Tat Lee and Aaron Sidford. Efficient accelerated coordinate descent methods and faster algorithms for solving linear systems. In *FOCS*, pages 147–156. IEEE, 2013.

[22] Michael W. Mahoney, Satish Rao, Di Wang, and Peng Zhang. Approximating the solution to mixed packing and covering lps in parallel $\widetilde{O}(\epsilon^{-3})$ time. In *ICALP*, 2016.

[23] H. Brendan McMahan and Matthew Streeter. Adaptive Bound Optimization for Online Convex Optimization. In *COLT*, 2010.

[24] Arkadi Nemirovsky and David Yudin. *Problem complexity and method efficiency in optimization*. Nauka Publishers, Moscow (in Russian), 1978. John Wiley, New York (in English) 1983.

[25] Yurii Nesterov. A method of solving a convex programming problem with convergence rate $O(1/k^2)$. In *Doklady AN SSSR (translated as Soviet Mathematics Doklady)*, volume 269, pages 543–547, 1983.

[26] Yurii Nesterov. *Introductory Lectures on Convex Programming Volume: A Basic course*, volume I. Kluwer Academic Publishers, 2004.

[27] Yurii Nesterov. Smooth minimization of non-smooth functions. *Mathematical Programming*, 103(1):127–152, December 2005.

[28] Yurii Nesterov. Primal-dual subgradient methods for convex problems. *Mathematical Programming*, 120(1):221–259, June 2007.

[29] Yurii Nesterov. Accelerating the cubic regularization of newton's method on convex problems. *Mathematical Programming*, 112(1):159–181, 2008.

[30] Yurii Nesterov. Gradient methods for minimizing composite functions. *Mathematical Programming*, 140(1):125–161, 2013.

[31] Yurii Nesterov. Universal gradient methods for convex optimization problems. *Mathematical Programming*, May 2014.

[32] Brendan O'Donoghue and Emmanuel Candès. Adaptive Restart for Accelerated Gradient Schemes. *Foundations of Computational Mathematics*, July 2013.

[33] Lorenzo Orecchia, Sushant Sachdeva, and Nisheeth K. Vishnoi. Approximating the exponential, the lanczos method and an $\widetilde{O}(m)$-time spectral algorithm for balanced separator. In *STOC '12*. ACM Press, November 2012.

[34] Serge A. Plotkin, David B. Shmoys, and Éva Tardos. Fast Approximation Algorithms for Fractional Packing and Covering Problems. *Mathematics of Operations Research*, 20(2):257–301, May 1995.

[35] Ankan Saha, S. V. N. Vishwanathan, and Xinhua Zhang. New Approximation Algorithms for Minimum Enclosing Convex Shapes. In *SODA*, pages 1146–1160, September 2011.

[36] Shai Shalev-Shwartz. Online Learning and Online Convex Optimization. *Foundations and Trends in Machine Learning*, 4(2):107–194, 2012.

[37] Shai Shalev-Shwartz and Yoram Singer. Logarithmic regret algorithms for strongly convex repeated games. Technical report, The Hebrew University, 2007.

[38] Shai Shalev-Shwartz and Tong Zhang. Accelerated Mini-Batch Stochastic Dual Coordinate Ascent. In *NIPS*, pages 1–17, May 2013.

[39] Shai Shalev-Shwartz and Tong Zhang. Accelerated Proximal Stochastic Dual Coordinate Ascent for Regularized Loss Minimization. In *ICML*, pages 64–72, 2014.

[40] Ohad Shamir and Tong Zhang. Stochastic Gradient Descent for Non-smooth Optimization:

Convergence Results and Optimal Averaging Schemes. In *Proceedings of the 30th International Conference on Machine Learning - ICML '13*, volume 28, 2013.

[41] Weijie Su, Stephen Boyd, and Emmanuel Candes. A differential equation for modeling nesterovs accelerated gradient method: Theory and insights. In *Advances in Neural Information Processing Systems*, pages 2510–2518, 2014.

[42] Di Wang, Michael W. Mahoney, Nishanth Mohan, and Satish Rao. Faster parallel solver for positive linear programs via dynamically-bucketed selective coordinate descent. *ArXiv e-prints*, abs/1511.06468, November 2015.

[43] Di Wang, Satish Rao, and Michael W. Mahoney. Unified acceleration method for packing and covering problems via diameter reduction. In *ICALP*, 2016.

[44] Lin Xiao. Dual averaging method for regularized stochastic learning and online optimization. *The Journal of Machine Learning Research*, 11:2543–2596, 2010.